\documentclass[10pt, twocolumn, notitlepage, superscriptaddress, prb, longbibliography]{revtex4-2}

\usepackage{here}
\usepackage{amsmath}         
\usepackage{graphicx}
\usepackage{braket}         
\usepackage{lettrine}
\usepackage{xcolor}
\usepackage{times}
\usepackage{hhline}
\usepackage{tabularx}
\usepackage{calc}
\usepackage{soul}

\usepackage{dcolumn}
\usepackage{bm}
\usepackage{float} 
\usepackage{subfigure}

\usepackage[colorlinks=true,linkcolor=blue, citecolor=blue, urlcolor=blue]{hyperref}%

\begin{document}

\preprint{APS/123-QED}

\title{Dzyaloshinskii-Moriya interaction from unquenched orbital angular momentum}

\author{Runze Chen}
\affiliation{%
 Fert Beijing Institute, School of Integrated Circuit Science and Engineering, Beijing Advanced Innovation Center for Big Data and Brain Computing, Beihang University, Beijing 100191, China
}%
\affiliation{
 Peter Grünberg Institut and Institute for Advanced Simulation, Forschungszentrum Jülich and JARA, 52425 Jülich, Germany
}%

\author{Dongwook Go}
\affiliation{
 Peter Grünberg Institut and Institute for Advanced Simulation, Forschungszentrum Jülich and JARA, 52425 Jülich, Germany
}%
\affiliation{
 Institute of Physics, Johannes Gutenberg University Mainz, 55099 Mainz, Germany
}%

\author{Stefan Blügel}
\affiliation{
 Peter Grünberg Institut and Institute for Advanced Simulation, Forschungszentrum Jülich and JARA, 52425 Jülich, Germany
}%

\author{Weisheng Zhao}%
\email{weisheng.zhao@buaa.edu.cn}
\affiliation{%
 Fert Beijing Institute, School of Integrated Circuit Science and Engineering, Beijing Advanced Innovation Center for Big Data and Brain Computing, Beihang University, Beijing 100191, China
}%

\author{Yuriy Mokrousov}%
\email{y.mokrousov@fz-juelich.de}
\affiliation{
 Peter Grünberg Institut and Institute for Advanced Simulation, Forschungszentrum Jülich and JARA, 52425 Jülich, Germany
}%
\affiliation{
 Institute of Physics, Johannes Gutenberg University Mainz, 55099 Mainz, Germany
}%

\date{\today}

\begin{abstract}
Orbitronics is an emerging and fascinating field that explores the utilization of the orbital degree of freedom of electrons for information processing.
An increasing number of orbital  phenomena are being currently discovered, with spin-orbit coupling mediating the interplay between orbital and spin effects, thus providing a wealth of control mechanisms and device applications. In this context, the orbital analog of spin Dzyaloshinskii-Moriya interaction (DMI), i.e.~{\it orbital DMI}, deserves to be explored in depth, since it is believed to be capable of inducing chiral orbital structures. Here, we unveil the main features and microscopic mechanisms of the orbital DMI in a two-dimensional square lattice using a tight-binding model of ${t_{2 g}}$ orbitals in combination with the Berry phase theory. This approach allows us to investigate and transparently disentangle the  
role of inversion symmetry breaking, strength of orbital exchange interaction and spin-orbit coupling in shaping the properties of the orbital DMI. By scrutinizing the band-resolved contributions we are  able to understand the microscopic mechanisms and guiding principles behind the orbital DMI and its anisotropy in two dimensional magnetic materials, and uncover a fundamental relation between the orbital DMI and its spin counterpart, which is currently explored very intensively. 
The insights gained from our work contribute to advancing our knowledge of orbital-related effects and their potential applications in spintronics, 
providing a path for future research in the field of chiral orbitronics.
\end{abstract}

\maketitle


\section{\label{sec:level1}INTRODUCTION}

In condensed matter physics, a new and promising area of research known as orbitronics has emerged in recent years. Orbitronics focuses on utilizing the orbital degree of freedom as a means of transmitting and manipulating information in solid-state devices~\cite{1,2,3}. Recent investigations have revealed intriguing possibilities within this realm, demonstrating that an orbital current, resulting from the movement of electrons possessing finite orbital angular momentum, can be generated in diverse range of materials, despite the presence of orbital quenching in the ground state ~\cite{4,5,6,7,8,9,10,11}. The fundamental principle behind orbitronics lies in the precise control of orbital dynamics and information transport and subsequent manipulation of spin transport and magnetization dynamics through spin-orbit coupling (SOC). 
This demonstrates the potential of utilizing the orbital current as an alternative to conventional spin current in the field of spintronics~\cite{12,13,14,15,16,17,18,19}. Moreover, the orbital degree of freedom serves as an intermediary, facilitating the appearance of diverse spin phenomena by mediating the coupling between magnetic moments and the lattice, thus triggering~e.g.~the $\mathbf{k}$-dependent spin splitting through structural inversion symmetry breaking and magnetization dynamics in an applied external electric field~\cite{4,20}. Correspondingly, in the past years the community witnessed a considerable interest in exploring the fundamental relation between various spin phenomena and their orbital analogs, such as orbital Rashba effect, orbital Hall effect and orbital torque~\cite{1,4,5,12,13,21,22,23,20,24,25,26,27,28,29,30,31,32,33,34,35,36,37,38}. 

In the realm of chiral spin magnetism, so-called the Dzyaloshinskii-Moriya interaction (DMI) has attracted significant attention due to its pivotal role in the formation and stabilization of topological magnetic textures, including chiral domain walls and magnetic skyrmions~\cite{42,43,44,45}, which hold great promise as information carries in emerging memory-storage and neuromorphic computing~\cite{46,47}. The DMI is an antisymmetric spin exchange interaction that originates from the combination of SOC with structural inversion asymmetry~\cite{39,40,41}. Recently, a growing body of theoretical and experimental research has illuminated the pivotal role played by orbital degrees of freedom in the emergence and manipulation of the DMI, thereby igniting a surge of interest in exploring orbital facets in the DMI. 

Importantly, Yamamoto {\it et al.} theoretically investigated the intricate connection between the DMI and orbital moments in heavy metal/ferromagnetic metal structures, revealing a close correlation between the sign of the DMI and the induced orbital moments of heavy metal elements~\cite{48}. Meanwhile, density functional theory calculations by Belabbes {\it et al.} showed a correlation of the DMI with an electric dipole moment at the oxide/ferromagnetic metal interface~\cite{49}. The electric dipole moment arises from the charge transfer triggered by orbital hybridization at the interface between an oxide and a ferromagnet, tightly correlated with the interfacial orbital moment. It established a strong relationship among the DMI, interfacial orbital hybridization, and the interfacial orbital moment. Subsequently, Nembach {\it et al.} conducted experiments to substantiate the calculations by Belabbes {\it et al.}, revealing that the DMI and the spectroscopic splitting factor, which measures the orbital moment, are indeed correlated~\cite{50}. Furthermore, Zhu {\it et al.} experimentally investigated the correlation between interfacial orbital hybridization and DMI in heavy metal/ferromagnetic metal structures, providing conclusive evidence of the pivotal role played by orbital hybridization of magnetic interfaces in the determination and regulation of the DMI~\cite{51}. It has also been found that the DMI appears to exhibit a close correlation with the orbital moment anisotropy. Kim {\it et al.} experimentally unveiled that the DMI is governed by the orbital anisotropy, attributing its microscopic origin to the asymmetric charge distribution at the interface due to electron hoppings driven by the inversion symmetry breaking (ISB)~\cite{52}. Moreover, the asymmetric charge distribution results in chiral orbital angular momentum, which is then converted into spin canting via SOC, thereby explaining the emergence of the DMI. 

At the same time, while it is clear that the various aspects of orbital nature and orbital contributions to the {\it spin} DMI governing the energetics of {\it spin} canting deserve further efforts, the physics and properties of the {\it orbital} DMI, which reflects the energy changes due to chiral canting of {\it orbital} moments, have been practically unexplored so far.  
Katsnelson {\it et al.} has applied the method of infinitesimal rotations to compute the values of orbital part of the DMI in magnetic La$_2$CuO$_4$ from LDA+$U$+SOC calculations, finding that the orbital part, while being generally small, is by far larger than the spin contribution~\cite{PhysRevB.82.100403}. On the other hand, Kim and Han have considered the superexchange interaction in a multi-orbital tight-binding Hubbard model, finding a contribution which is proportional to a chiral product between neighboring orbital moments~\cite{53}. Identifying strong on-site correltaions and ISB as the origin of orbital version of DMI, they have demonstrated that the latter can dictate the formation of complex orbital textures~\cite{53}.

The orbital DMI and the resulting orbital chiral structures present a novel avenue for creating and manipulating chiral topological textures or angular momentum by controlling orbital properties, marking the orbital DMI as an object  deserving further in-depth investigation. Therefore, a comprehensive understanding of orbital DMI is imperative. However, a clear and simple physical picture of the orbital DMI, reaching into the realm of electronic structure models not necessarily rooted in strong correlations, is still elusive, which necessitates further clarification of the influence of various orbital parameters on the orbital DMI.

Here, we evaluate and analyze in depth the orbital DMI behavior in a two-dimensional square lattice based on a simple tight-binding model of ${t_{2 g}}$ orbitals. This model satisfies the requirement that the crystal field splitting in combination with ISB promote rich unquenched orbital magnetism, making it an optimal choice to extract the orbital-related effects~\cite{28}. We resort to the Berry phase theory of orbital DMI, which is obtained as a direct extension of the successful methodology to compute the spin DMI, developed in the past~\cite{57}. As one of the key variables in this method, we introduce the orbital exchange coupling term in the Hamiltonian, discussing its nature and physical origins, assessing the orbital DMI as the energy response of the model to its spiralization, and relating it to the mixed orbital Berry curvature. We demonstrate that the orbital hybridization induced by ISB not only leads to chiral orbital angular momentum (OAM) structures in momentum space but also gives rise to prominent orbital DMI localized in specific regions of the Brillouin zone. Furthermore, we expore in which way the orbital exchange and ISB strengths impact the orbital DMI.
We find that the orbital DMI exhibits a strong anisotropy with respect to the orbital magnetization direction. We also show that even in the absence of SOC, the orbital DMI persists while the spin DMI vanishes, finding that non-relativistic by nature orbital DMI can dictate the behavior of spin DMI, consistent with the behavior of other orbital effects. With our work we thus contribute significantly to putting the orbital DMI physics onto the rails of material engineering in wider classes of two-dimensional metallic surfaces, interfaces and heterostructures, opening new perspectives for experimental realization of chiral orbital textures.

\section{METHODS}

\subsection{Motivation and physical setups}

Before proceeding further with a detailed description of our approach, it is necessary to elucidate the motivation behind our treatment of OAM contribution to the DMI. Our starting point is the non-relativistic non-magnetic Hamiltonian of an orbitally complex system, represented by a ${t_{2 g}}$ orbital model on a square lattice with broken inversion symmetry.  Starting with this model, our next step is to break the time-reversal symmetry by allowing unquenched local orbital moments, which serve as the primary objects whose canting energetics we are to study. The conventional way to achieve this would be to allow for spin exchange interaction which, when combined with an effect of SOC, results in orbital magnetism. However, here, we choose a different path and explicitly include the effect of the orbital exchange field into consideration, which couples directly to the OAM, see Eq.~\eqref{eq:orbital_exchange} below, and gives rise to unquenched orbital magnetism without resorting to spin magnetism and SOC. In turn, one can perceive the effect of the orbital exchange field as a combined effect of the two latter phenomena, projected onto the orbital subspace only, which allows for a transparent way to perturb the directions of orbital moments by tracking the energetics of the orbital sub-system. 

When thinking of a physical situation in which the effect of the orbital exchange field can be most easily singled out, we can imagine a weakly coupled bilayer type of system, where a nonmagnetic layer develops orbital magnetism via the SOC and the nonlocal exchange coupling with the second layer that is strongly magnetic, e.g. 3$d$ overlayer separated by a spacer from a 5$d$ substrate ~\cite{54,55,56}. In the latter case, considering that we can freely impose magnetic order in the 3$d$ metal, this will be translated into corresponding orbital order via the effect of SOC and effective orbital field, the energetics of which we can study separately assuming that it would be possible to neglect the spin contributions arising in the 5$d$ layer itself due to smallness of corresponding induced spin moments. 

Secondly, we can consider a situation in which the shell of orbitals which drive spin magnetism is separated {\it in energy} from a nominally spin-degenerate but orbitally active shell of states residing around the Fermi level, such as the case e.g. for materials incorporating $4f$-ions~\cite{zeer2023promoting}. Here, again, the SOC-imposed contribution to (e.g. $p$-$d$-$f$) hybridization between the two shells would results in the presence of an effective orbital field. In the latter two cases, for clear identification of the orbital DMI, it seems to be  particularly promising to consider antiferromagnetic bilayers and $f$-based compounds, where the net effect of spin splitting on the orbital states can be vanishingly small, and they can be considered practically spin degenerate~\cite{zeer2023promoting}. 

Finally, we can turn to a situation of a non-magnetic thin film exposed to a spatially varying external magnetic field which drives a direct interaction with orbital moments in the substrate of the Zeeman type $-$ i.e. orbital Zeeman effect $-$ as given by Eq.~\eqref{eq:orbital_exchange}. The latter case has been considered recently in a situation of orbital pumping, where magnetic field dynamics drives a prominent orbital response~\cite{han2023theory,go2023orbital}.

In the following, we provide the details of the ${t_{2 g}}$ orbital model and the derivations of the expression of orbital DMI.

\subsection{${t_{2 g}}$ orbitals on a two-dimensional square lattice}

We employ a tight-binding model description of a simple two-dimensional square lattice with ${t_{2 g}}$ orbitals (${d_{x y}, d_{y z}, d_{z x}}$) on each site. 
We operate in terms of Bloch waves 
$
e^{i \mathbf{k} \cdot \mathbf{r}}\left|\varphi_{n \mathbf{k}}\right\rangle=\sum_{\mathbf{R}} e^{i \mathbf{k} \cdot \mathbf{R}}\left|\phi_{n \mathbf{R}}\right\rangle
$ 
as basis, where $\phi_{n \mathbf{R}}$ is the localized state at the Bravais lattice $\mathbf{R}$ with the orbital character $n$, $n=d_{x y}, d_{y z}, d_{z x}$. The spinless tight-binding Hamiltonian in $\mathbf{k}$-space is written as
\begin{equation}
H_{\text {tot}}(\mathbf{k})=H_{\text {kin}}(\mathbf{k})+H_{\text {exc}}^{\mathrm{L}}, 
\end{equation}
where $H_{\text {kin}}(\mathbf{k})$ is the kinetic part of the Hamiltonian arising from hoppings and on-site energies, and $H_{\mathrm{exc}}^{\mathrm{L}}$ describes orbital exchange interaction. $H_{\text {kin}}(\mathbf{k})$ is independent of the spin and its nonzero matrix elements are
\begin{subequations}
\begin{align}
& \left\langle\varphi_{d_{x y} \mathbf{k}}\left|H_{\mathrm{kin}}\right| \varphi_{d_{x y} \mathbf{k}}\right\rangle=E_{\|}-2 t_\delta\left[\cos \left(k_x a\right)+\cos \left(k_y a\right)\right], \\
& \left\langle\varphi_{d_{y z} \mathbf{k}}\left|H_{\mathrm{kin}}\right| \varphi_{d_{y z} \mathbf{k}}\right\rangle=E_{\perp}-2 t_\pi \cos \left(k_x a\right)-2 t_\delta \cos \left(k_y a\right), \\
& \left\langle\varphi_{d_{z x} \mathbf{k}}\left|H_{\mathrm{kin}}\right| \varphi_{d_{z x} \mathbf{k}}\right\rangle=E_{\perp}-2 t_\pi \cos \left(k_y a\right)-2 t_\delta \cos \left(k_x a\right), \\
& \left\langle\varphi_{d_{x y \mathbf{k}}}\left|H_{\mathrm{kin}}\right| \varphi_{d_{y z} \mathbf{k}}\right\rangle=2 i \gamma \sin \left(k_x a\right), \\
&  \left\langle\varphi_{d_{x y} \mathbf{k}}\left|H_{\mathrm{kin}}\right| \varphi_{d_{z x} \mathbf{k}}\right\rangle=2 i \gamma \sin \left(k_y a\right).
\end{align}
\end{subequations}
Here, $E_{\|}$ and $E_{\perp}$ are on-site energies for the in-plane ($d_{x y}$) and out-of-plane ($d_{y z}$ and $d_{z x}$) orbitals, respectively, and $t_\pi$ and $t_\delta$ are the nearest neighbor hopping amplitudes between $t_{2 g}$ orbitals via the $\pi$ and $\delta$ bondings, respectively. The inversion symmetry breaking at the surface is equivalent to the potential gradient along the surface-normal direction ($z$), which promotes hybridization between $d_{x y}$ and $d_{y z}$, as well as  $d_{x y}$ and $d_{z x}$ states, with hopping integral $\gamma$ characterizing the strength of the symmetry breaking. The orbital exchange interaction is written as
\begin{equation}
\label{eq:orbital_exchange}
H_{\mathrm{exc}}^{\mathrm{L}}=\frac{J_L}{\hbar} \hat{\mathbf{M}}^{\mathrm{L}} \cdot \mathbf{L},
\end{equation}
where $\mathbf{L}$ is the atomic OAM operator, $\hat{\mathbf{M}}^{\mathrm{L}}$ denotes the direction of the orbital exchange field, and $J_L$ denotes the coupling strength of the orbital exchange interaction. For $t_{2 g}$ orbitals, $\boldsymbol{L}=\left(L_x, L_y, L_z\right)$ becomes
\begin{subequations}
\begin{align}
& L_x=\hbar\left(\begin{array}{ccc}
0 & 0 & -i \\
0 & 0 & 0 \\
i & 0 & 0
\end{array}\right), \\
& L_y=\hbar\left(\begin{array}{ccc}
0 & i & 0 \\
-i & 0 & 0 \\
0 & 0 & 0
\end{array}\right), \\
& L_z=\hbar\left(\begin{array}{ccc}
0 & 0 & 0 \\
0 & 0 & i \\
0 & -i & 0
\end{array}\right) \text {, }
\end{align}
\end{subequations}
in the matrix representation using the basis states $\varphi_{d_{x y} \mathbf{k}}, \varphi_{d_{y z} \mathbf{k}}$, and $\varphi_{d_{z x} \mathbf{k}}$. 

A spinfull model Hamiltonian is constructed by adding SOC $H_{\rm {soc}}$, and spin exchange interaction $H_{\mathrm{exc}}^{\mathrm{S}}$. The SOC is given by
\begin{equation}
H_{\text {soc}}=\frac{2 \lambda_{\text {soc}}}{\hbar^2} \mathbf{L} \cdot \mathbf{S} \text {, }
\end{equation}
where $\mathbf{S}$ is the spin angular momentum operator and $\lambda_{\text {soc}}$ is the SOC strength. The spin exchange interaction is written as
\begin{equation}
H_{\mathrm{exc}}^{\mathrm{S}}=\frac{J_S}{\hbar} \hat{\mathbf{M}}^{\mathrm{S}} \cdot \mathbf{S},
\end{equation}
where $\hat{\mathbf{M}}^{\mathbf{S}}$ denotes the direction of the spin exchange field, and $J_S$ denotes the spin exchange interaction strength. The spin angular momentum operator is represented by the vector of the Pauli matrices $\boldsymbol{\sigma}=(\sigma_x,\ \sigma_y,\ \sigma_z)$ within each orbital
\begin{equation}
\left\langle\varphi_{n \sigma \mathbf{k}}|\mathbf{S}| \varphi_{n \sigma^{\prime} \mathbf{k}}\right\rangle=\frac{\hbar}{2}[\boldsymbol{\sigma}]_{\sigma \sigma^{\prime}}
\text {, }
\end{equation}
where $\sigma,\sigma'$ are spin indices (up or down). 

The values of the model parameters used in the calculation are $E_{\|}=1.6, E_{\perp}=1.0, t_\pi=0.5, t_\sigma=0.1, \gamma=0.02, \lambda_{\text {soc }}=$ $0.04, J_L=1.0, J_S=1.0$, all in unit of eV. The lattice constant is set $a=1$ \AA. All parameters are set as above unless specified otherwise.

\subsection{Berry phase expression for the orbital DMI}

We define the orbital DMI as the energy contribution to the free energy functional by the spiralization of the orbital exchange field $\hat{\mathbf{M}}^{\mathrm{L}}$,
\begin{equation}
\mathcal{F}^\mathrm{L}_\mathrm{DMI}=
\int d^2 r
\sum_{i j} D_{i j}^{\mathrm{L}} \hat{\mathbf{r}}_i \cdot\left(\hat{\mathbf{M}}^{\mathrm{L}} \times \partial_{r_j} \hat{\mathbf{M}}^{\mathrm{L}}\right)
\text {, }
\label{DMI-F}
\end{equation}
where $\hat{\mathbf{r}}_i$ is the unit vector in the $i$ direction, and $\partial_{r_j}$ is the partial derivative in the $j$ direction ($i,j=x,y,z$). We derive a microscopic expression for the orbital DMI coefficient $D_{i j}^{\mathrm{L}}$ by means of the Berry phase theory, which has been applied in the past to calculate the spin DMI~\cite{57,58,59}. In this method,  $D_{i j}^{\mathrm{L}}$ can be quantified by expanding the free energy in terms of small spatial gradients of orbital magnetization direction within quantum mechanical perturbation theory.

Given the orbital exchange interaction in the spinless Hamiltonian in Eq.~\eqref{eq:orbital_exchange}, we define the torque operator on the orbital exchange field due to the OAM by 
\begin{equation}
\boldsymbol{\mathcal{T}}^{\mathrm{L}}=-\frac{1}{i\hbar}[\mathbf{L},\ H_\mathrm{exc}^\mathrm{L}]=\frac{J_L}{\hbar} \hat{\mathbf{M}}^\mathrm{L}\times\mathbf{L}.
\end{equation}
Following the detailed derivation in Ref.~\cite{58}, the first-order perturbation by the spiralization of $\hat{\mathbf{M}}^\mathrm{L}$ starting from a collinear configuration of $\hat{\mathbf{M}}^\mathrm{L}$ gives rise the the following expression in terms of electronic states
\begin{eqnarray}
D_{i j}^{\mathrm{L}}
&=&\int \frac{d^2k}{(2\pi)^2} \sum_{n}\Big[ f\left(\varepsilon_{n\mathbf{k}}\right) A_{i j}^n(\mathbf{k})
\nonumber
\\
& &  +\frac{1}{\beta} \ln \left(1+e^{-\beta\left(\varepsilon_{n\mathbf{k}}-\mu\right)}\right) B_{i j}^n(\mathbf{k})\Big]
\text {, }
\label{D}
\end{eqnarray}
where $f\left(\varepsilon_{n\mathbf{k}}\right)=1/[1+e^{\beta(\varepsilon_{n\mathbf{k}}-\mu)}]$ is the Fermi-Dirac distribution function for the unperturbed band energy $\varepsilon_{\mathrm{k} n}$, $\mu$ is the electrochemical potential, and $\beta=1/k_\mathrm{B}T$ for the Boltzmann constant $\mathbf{k}_\mathrm{B}$ and the temperature $T$. The quantities $A_{i j}^n (\mathbf{k})$ and $B_{i j}^n (\mathbf{k})$ are given by the correlation between the orbital torque operator and the velocity operator, 
\begin{equation}
A_{i j}^n(\mathbf{k})=-\hbar  \sum_{m \neq n} \frac{
\mathrm{Im}
\left[
\left\langle u_{n\boldsymbol{k}}\left|\mathcal{T}_i^L\right| u_{m\boldsymbol{k}}\right\rangle\left\langle u_{m\boldsymbol{k}}\left| v_j\right| u_{n\boldsymbol{k}}\right\rangle
\right]
}{\varepsilon_{n\boldsymbol{k}}-\varepsilon_{m\boldsymbol{k}}}
\text {, }
\end{equation}
and
\begin{equation}
B_{i j}^n(\mathbf{k})=-2 \hbar  \sum_{m \neq n} \frac{
\mathrm{Im}
\left[
\left\langle u_{n\mathbf{k}}\left|\mathcal{T}_i^L\right| u_{m\mathbf{k}}\right\rangle\left\langle u_{m\mathbf{k}}\left| v_j\right| u_{n\mathbf{k}}\right\rangle
\right]
}{\left(\varepsilon_{n\mathbf{k}}-\varepsilon_{m\mathbf{k}}\right)^2}
\text {, }
\end{equation}
where $u_{n\mathbf{k}}$ denotes the lattice-periodic part of the unperturbed Bloch wave function of band $n$ at $\mathbf{k}$ and $v_j=\partial_{k_j} H_\mathrm{kin}/\hbar$ is the $j$-th component of the velocity operator.

At zero temperature Eq.~(\ref{D}) becomes
\begin{equation}
D_{i j}^{\mathrm{L}}=\int \frac{d^2k}{(2\pi)^2} \sum_{n} f (\varepsilon_{n\mathbf{k}}) \left[A_{i j}^n(\mathbf{k})-\left(\varepsilon_{n\mathbf{k}}-\mu\right) B_{i j}^n(\mathbf{k})\right]
\text {. }
\label{eq:DMI_zero_T}
\end{equation}
By utilizing $\boldsymbol{T}^\mathrm{L}=\hat{\mathbf{M}}^\mathrm{L}\times \partial_{\hat{\mathbf{M}}^\mathrm{L}} H_\mathrm{tot}$, Eq.~\eqref{eq:DMI_zero_T} becomes
\begin{eqnarray}
D_{i j}^{\mathrm{L}}
&=&
\int \frac{d^2k}{(2\pi)^2} \sum_{n} f (\varepsilon_{n\mathbf{k}}) 
\hat{\mathbf{r}}_i \cdot 
\label{eq:DMI_zero}
\\
& & 
\nonumber
\mathrm{Im}
\left[ 
\hat{\mathbf{M}}^\mathrm{L} \times 
\bra{\partial_{\mathbf{M}^\mathrm{L}} u_{n\mathbf{k}}} 
\left(
H_\mathrm{tot} + \varepsilon_{n\mathbf{k}} - 2\mu 
\right)
\ket{\partial_{k_j} u_{n\mathbf{k}}}
\right]
\text {. }
\end{eqnarray}
Note the similarity of this expression to the mixed Berry curvature
\begin{equation}
\Omega_{M_i^\mathrm{L} k_j}
=
-2 \hat{\mathbf{r}}_i \cdot \operatorname{Im} 
\left[ 
\hat{\mathbf{M}}^\mathrm{L} \times 
\braket{\partial_{\mathbf{M}^\mathrm{L}} u_{n\mathbf{k}}
|
\partial_{k_j} u_{n\mathbf{k}}}
\right].
\label{Omega}
\end{equation}
Up to date, the properties of orbital mixed Berry curvature in solid state systems have been not explored. According to Eq.~\eqref{eq:DMI_zero}, the orbital spiralization, in analogy to spin spiralization and modern theory of orbital magnetization~\cite{60,61,62}, comprises two terms: one, ``itinerant" in the language of orbital magnetization, proportional to the orbital mixed Berry curvature and associated with the $B_{ij}^n$-part in Eq.~(\ref{eq:DMI_zero_T}), and another one, which should be treated as the ``self-rotation" contribution, driven by the $A_{ij}^k$-part in Eq.~(\ref{eq:DMI_zero_T}). While the second part of the orbital DMI is arising from the local breaking of inversion symmetry, the itinerant mixed Berry curvature part can be attributed to the breaking of inversion symmetry at the boundary of a finite sample taken to the thermodynamic limit.

In considering the spin, as mentioned above, the Hamiltonian for spinfull systems necessitates the inclusion of both spin-orbit coupling and spin exchange interaction. In this case, the spin torque operator can be represented as $\boldsymbol{\mathcal{T}}^{\mathrm{S}}=(J_S/\hbar) \hat{\mathbf{M}}^\mathrm{S}\times\mathbf{S}$. From this, we can obtain the spin spiralization $D_{i j}^{\mathrm{S}}$ through the above equations by replacing the orbital exchange field $\hat{\mathbf{M}}^\mathrm{L}$ by the spin exchange field $\hat{\mathbf{M}}^\mathrm{S}$.

\section{RESULTS}

\subsection{Orbital DMI by orbital Rashba effect in $\mathbf{k}$-space}

First, let us unravel the emergence of an intrinsic orbital texture of $t_{2 g}$ orbitals without considering the effect of SOC and orbital/spin exchange interaction. In Figs.~1(a)-(b) we present the band structure of the model along the high symmetry directions and near $\Gamma$, respectively. We can see clearly in Fig.~1(b) the effect of orbital hybridization mediated by inversion symmetry breaking, inducing avoided crossings of the bands with different orbital characters. In Figs.~1(c)-(e), we show the distribution of the expectation value of OAM in $\mathbf{k}$-space near the $\Gamma$ point for the bands $E_{1, \mathbf{k}}$, $E_{2, \mathbf{k}}$, and $E_{3, \mathbf{k}}$, which are labeled in the increasing order of energy. We observe that the OAM exhibits a chiral texture around the $\Gamma$ point, specifically, of clockwise, counter-clockwise, and zero character around the $\Gamma$ point, respectively. The chiral behavior of the $t_{2 g}$ orbital textures is driven by the orbital Rashba effect of the form $\hat{z} \cdot(\mathbf{k} \times \mathbf{L})$, which is consistent with previous studies~\cite{28}. Investigating the influence of SOC reveals that the spin Rashba effect plays a crucial role in driving the spin texture's chirality appearing on top of the orbital distribution. As a result, the spin textures are closely tied to the local orientation of the OAM in the Brillouin zone, manifesting as either parallel or antiparallel alignment with the underlying OAM (See supplementary section S1~\cite{63}).

\begin{figure}[t!] 
\centering 
\includegraphics[width=0.48\textwidth]{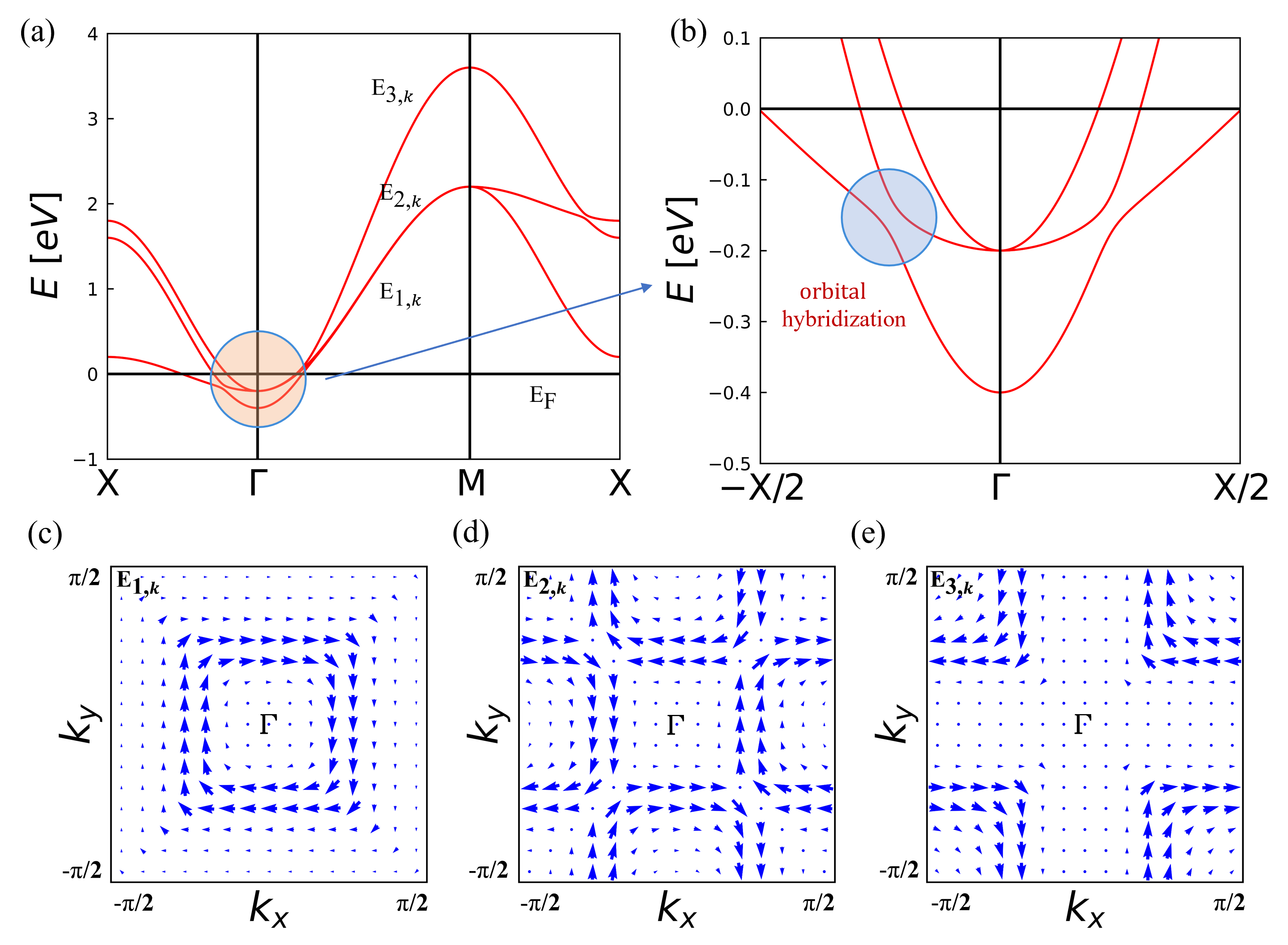}
\caption{ {\bf Orbital texture of the spinless model.} (a) The bandstructure of the $t_{2 g}$ model along the high symmetry directions for two-dimensional square lattice calculated without SOC and spin/orbital exchange interaction. (b) The band structure near $\Gamma$ corresponding to the region marked with an orange circle in (a). The blue circle marks the region of strong orbital hybridization mediated by inversion symmetry breaking. (c)-(e) Band-resolved orbital textures in the Brillouin zone, where 
$E_{1, {\bf k}}$, $E_{2, {\bf k}}$, and  $E_{3, {\bf k}}$ denote the different bands as indicated in (a). The direction of orbital angular momentum is shown with an arrow with a length corresponding to its magnitude (in arbitrary units).} 
\label{Fig.1} 
\end{figure}

\begin{figure}[t!] 
\centering 
\includegraphics[width=0.55\textwidth]{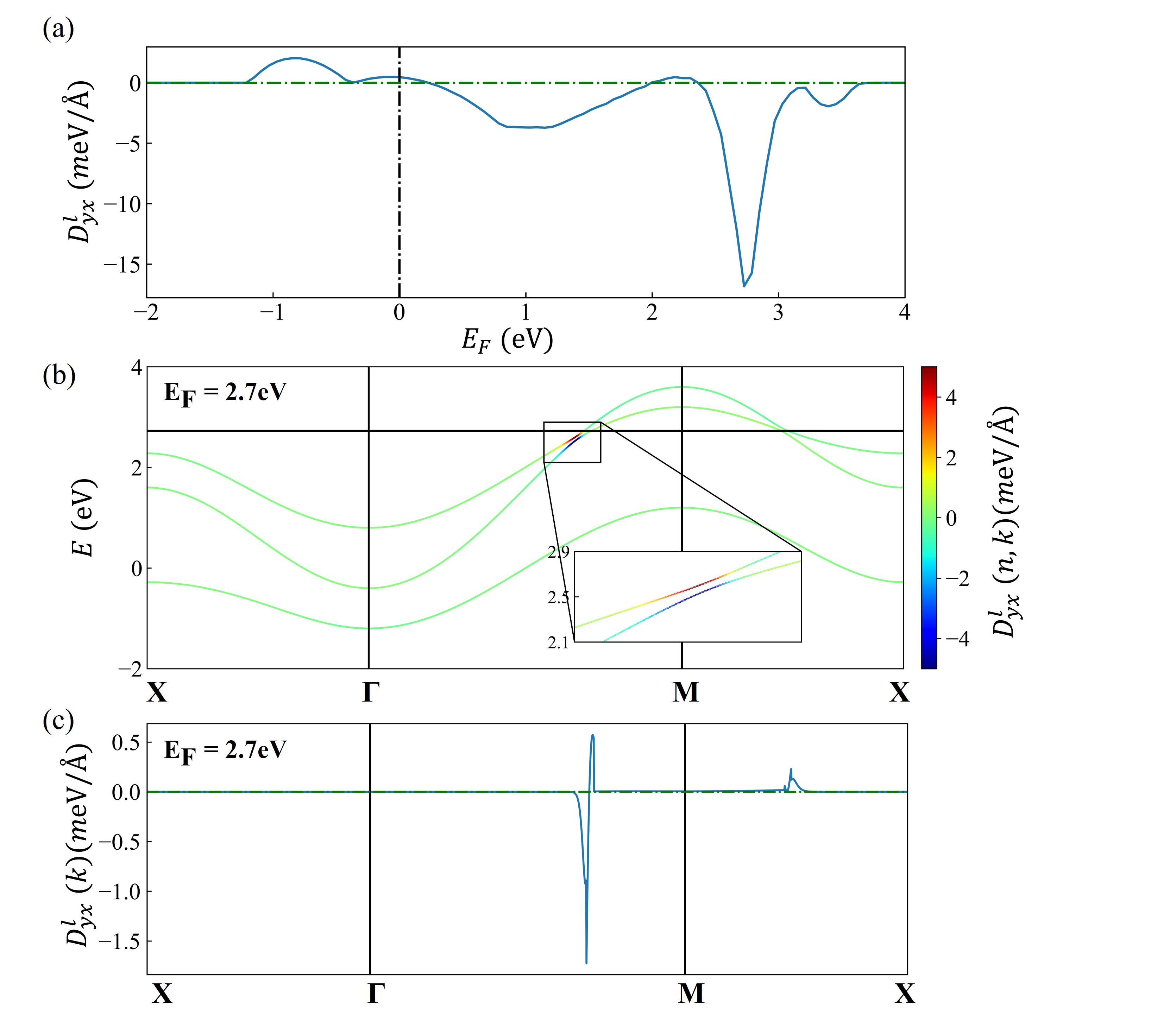} 
\caption{{\bf Orbital DMI without SOC}. (a) The orbital DMI variation with respect to the position of Fermi energy without SOC. (b) The band-projected orbital DMI along the high symmetry path in the Brillouin zone at $E_F=2.7$\,eV. The color marks the value of orbital DMI projected on each band. The projected orbital DMI is pronounced in the vicinity of the band avoided crossing between $\Gamma$ and $\mathrm{M}$ points. (c) The $\mathbf{k}$-resolved orbital DMI along the high symmetry path in the Brillouin zone at $E_F=2.7$\,eV.} 
\label{Fig.2} 
\end{figure}

Next, we consider the effect of the orbital Zeeman effect $H_{\mathrm{exc}}^{\mathrm{L}}$ due to the orbital exchange field $\hat{\mathbf{M}}^{\mathrm{L}}$ (keeping the value of $J_L$ at 1.0\,eV). In the two-dimensional square lattice with an out-of-plane inversion symmetry breaking, the orbital DMI spiralization is described by an antisymmetric tensor due to the four-fold rotational symmetry around the axis normal to the film, which amounts to $D_{x x}^{\mathrm{L}}=D_{y y}^{\mathrm{L}}=0$ and $D_{y x}^{\mathrm{L}}=-D_{x y}^{\mathrm{L}}$. We compute the magnitude of $D_{y x}^{\mathrm{L}}$ and present its magnitude in Fig.~2(a) as a function of band filling as given by the position of the Fermi level $E_F$, without SOC and with the orbital exchange field pointing out of the plane. The computed oscillatory behavior is strongly reminiscent of the Fermi energy dependence of the spin DMI typical for thin films of transition metals~\cite{59}. Remarkably, we observe that the magnitude of orbital DMI spiralization $D_{y x}^{\mathrm{L}}$ within our model reaches a large value of $16.8$\,meV/\AA\, for $E_F=2.7$\,eV, which is comparable to the theoretical magnitude of spin DMI in typical transition-metal systems, such as~e.g.~Co/Pt, Co/Ir and Co/Au thin films~\cite{58,59}. This finding serves as compelling evidence that significant orbital DMI can manifest in a system lacking SOC but with inversion symmetry breaking when this system possesses a strong orbital exchange interaction. 

More insight can be gained from a band-projected analysis of orbital DMI at $E_F=2.7$\,eV, as shown in Fig.~2(b), where the color marks the value of projected orbital DMI onto each band. We observe that significant orbital DMI can be observed predominantly in the vicinity of the avoided band crossing between $\Gamma$ and $\mathrm{M}$ points in the Brillouin zone. 
As shown in Fig.~2(c), the $\mathbf{k}$-resolved orbital DMI displays a very spiky behavior in this region, which corresponds to the region of orbital hybridization  associated with inversion symmetry breaking. At this Fermi energy, another significant contribution to the orbital DMI can be seen between M and X in the region of another band anti-crossing along that path. This underlines the crucial importance of the band-crossing points for achieving large values of the orbital DMI in realistic systems.
\begin{figure}[t!] 
\centering 
\includegraphics[width=0.45\textwidth]{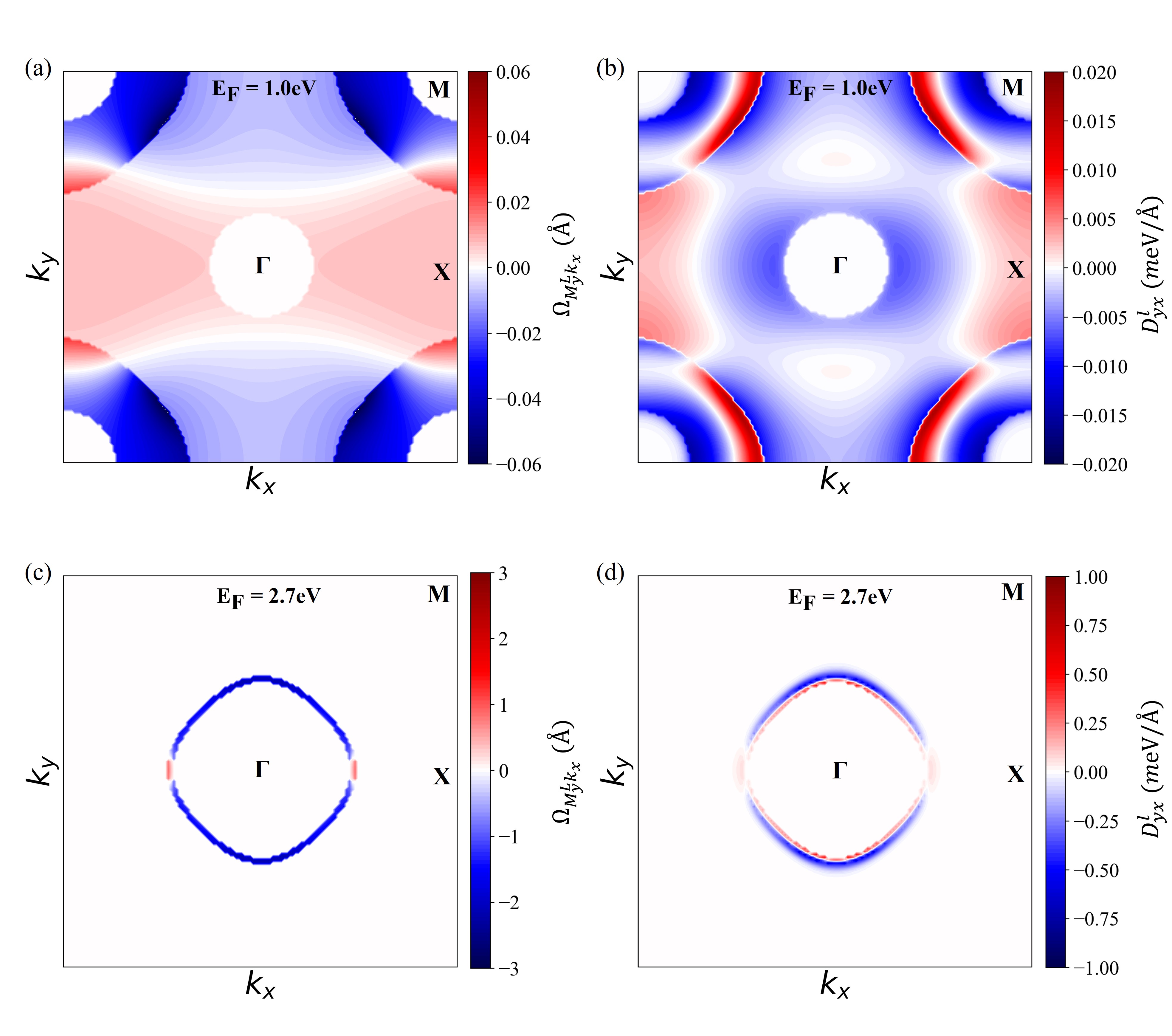}
\caption{{\bf Orbital mixed Berry curvature}. Distribution of the orbital mixed Berry curvature $\Omega_{M_y^\mathrm{L} k_x}$ (left) and the orbital Dzyaloshinskii-Moriya spiralization $D_{y x}^{\mathrm{L}}$ (right) of all occupied bands in the  Brillouin zone of the model for two Fermi energy values: (a)-(b) $E_F=1.0$\,eV and (c)-(d) $E_F=2.7$\,eV. All calculations are performed without considering SOC. The data are shown on a logarithmic scale, ${\rm sgn}(x)\log(1+|x|)$.} 
\label{Fig.3} 
\end{figure}

Given a close relation between the orbital DMI and orbital mixed Berry curvature, it is insightful to compare the two quantities directly. In Fig.~3 we plot the $\mathbf{k}$-resolved distribution of the orbital spiralization $D_{yx}^\mathrm{L}$ and of the mixed Berry curvature $\Omega_{M_y^\mathrm{L} k_x}$, which are summed over the occupied bands below $E_F=1.0$\,eV and $E_F=2.7$\,eV, respectively.  
At $E_F=1.0$\,eV, both $\Omega_{M_y^\mathrm{L} k_x}$ and $D_{y x}^{\mathrm{L}}$ exhibit complex  patterns with significant background contributions arising from broad areas of the Brillouin zone. Moreover, they show relatively small values throughout the entire Brillouin zone, consistent with the low values of the orbital DMI at this energy as visible in Fig.~1(a). In contrast, when $E_F=2.7$\,eV, both $\Omega_{M_y^\mathrm{L} k_x}$ and $D_{y x}^{\mathrm{L}}$ are sharply peaked in narrow regions of the Brillouin zone, corresponding to quasi-nodal lines arising due to near degenerate bands crossing the Fermi level. Clearly, the nodal lines serve as prominent sources of both orbital mixed Berry curvature and orbital DMI, with the two quantities being directly correlated in sign and magnitude.

Next, we explore the impact of SOC on orbital DMI. Namely, we perform calculations to examine the energy dependence, band dispersion, and $\mathbf{k}$-space distribution of orbital DMI while considering the presence of SOC. Remarkably, we find that the qualitative nature of orbital DMI remains unaffected by SOC, see supplementary 
sections S2 and S3~\cite{63}. Our findings align with previous studies on other orbital effects, such as the orbital Rashba effect and orbital Hall effect~\cite{4,20}, indicating that SOC does not play a decisive role in the emergence of orbital DMI, although it impacts its magnitude via the influence on the electronic structure details, see also discussion below. 

\subsection{Anatomy of orbital DMI}

\begin{figure}[t!] 
\centering 
\includegraphics[width=0.43\textwidth]{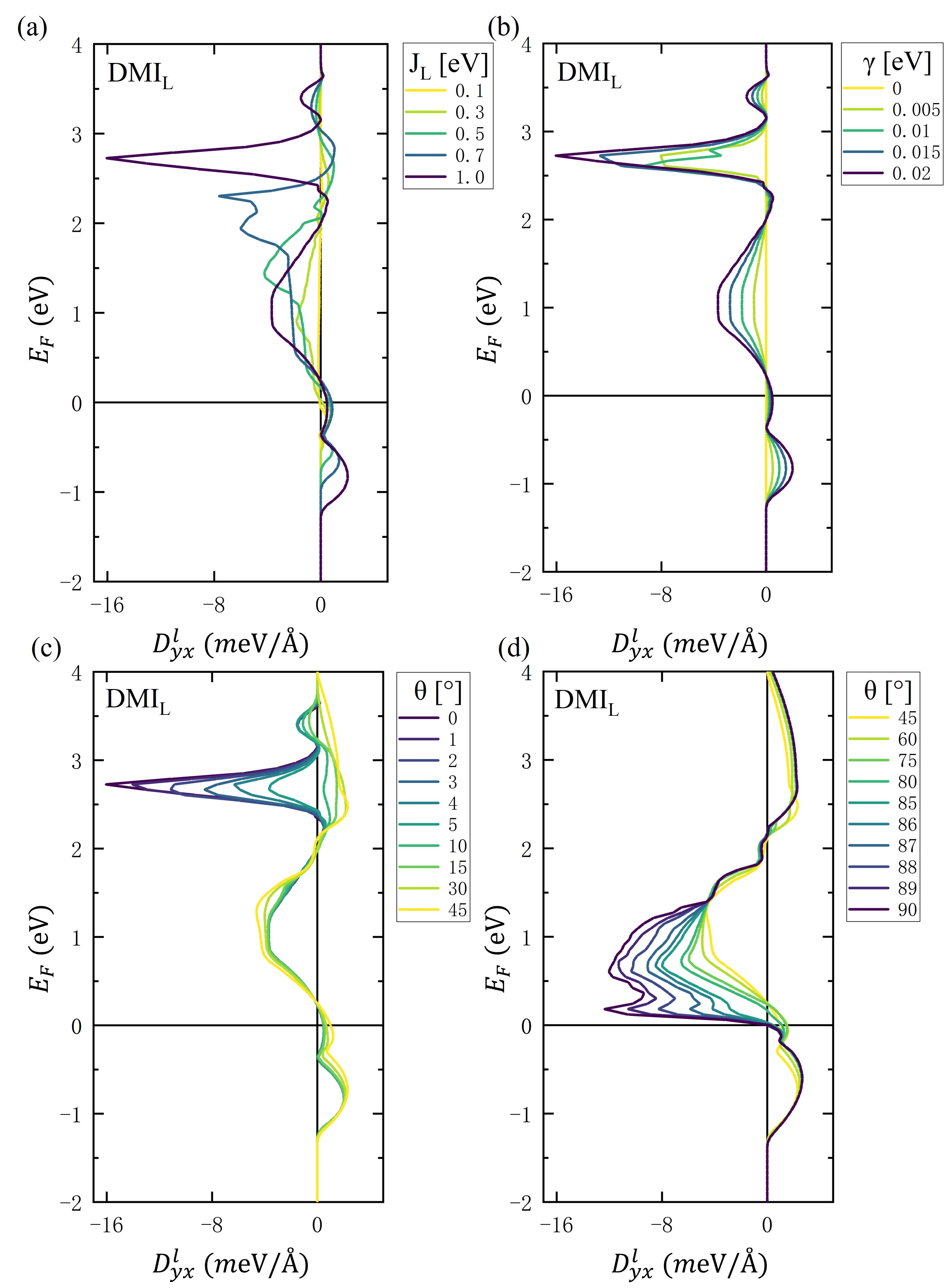} 
\caption{{\bf The anatomy of orbital DMI}. The orbital DMI $\left(D_{y x}^{\mathrm{L}}\right)$ as a function of the Fermi energy $\left(E_F\right)$ for (a) different orbital exchange coupling strengths $J_L$ and (b) inversion symmetry breaking strengths $\gamma$. The orbital DMI $\left(D_{y x}^{\mathrm{L}}\right)$ as a function of the Fermi energy $\left(E_F\right)$ for different angles $\theta$ of the orbital exchange field with the $z$-axis  in the ranges of (c) $0^{\circ}$ to $45^{\circ}$ and (d) $45^{\circ}$ to $90^{\circ}$. All calculations take into account the effect of SOC ($\lambda_{\text {soc }}=0.04$\,eV).} 
\label{Fig.4} 
\end{figure}

We analyze in great detail to reveal the anatomy of the orbital DMI. At this point, by keeping the SOC strength at a constant value of 0.04\,eV, we study the behavior of the orbital DMI in response to changing the key parameters of the system $-$ the strength of orbital exchange coupling $J_L$, the strength of inversion symmetry breaking $\gamma$, and the angle $\theta$ that the orbital exchange field makes with the $z$-axis $-$ presenting the results in Fig.~4.  
In Fig.~4(a), the orbital DMI $D_{y x}^{\mathrm{L}}$ is shown as a function of the Fermi energy $E_F$, for increasing orbital exchange coupling strength $J_L$. It is noteworthy that with increasing $J_L$, both the amplitude and position of the peak exhibit distinct deviations. Specifically, the height of the peak exhibits a positive correlation with $J_L$, whereas the peak position undergoes an upward shift with increasing $J_L$. This observation underscores the critical significance of orbital exchange coupling in determining the magnitude of the orbital DMI. When $J_L$ is sufficiently small, the orbital DMI assumes a significantly diminished value, aligning with the general scenario in practical materials where the orbital Zeeman effect is negligibly small and consequently conceals the manifestation of the orbital DMI. 

In order to gain a deeper understanding of the trend of orbital DMI with respect to the orbital exchange coupling strength $J_L$, we conduct band-projected as well as $\mathbf{k}$-resolved calculations of orbital DMI. In Fig.~5 we compare two cases $-$ $J_L=0.3$\,eV and $J_L$ $=0.7$\,eV $-$ with the Fermi energy set to the peak position for each case. It is evident from Figs.~5(a) and (b) that as $J_L$ increases,  the position of the band avoided crossing region moves up in energy in response to band dynamics. This is directly reflected in a shift of the peak in orbital DMI as seen in Fig.~4(a). Furthermore, by comparing Figs.~5(c) and (d), we observe an increase in the absolute value of $D_{y x}^{\mathrm{L}}(k)$ in the $\mathbf{k}$-region associated with orbital hybridization, resulting in an enhanced peak height in Fig.~4(a) with increasing $J_L$. The analysis for other values of $J_L$ in presented in the supplementary section S4~\cite{63}.

\begin{figure}[t!] 
\centering 
\includegraphics[width=0.48\textwidth]{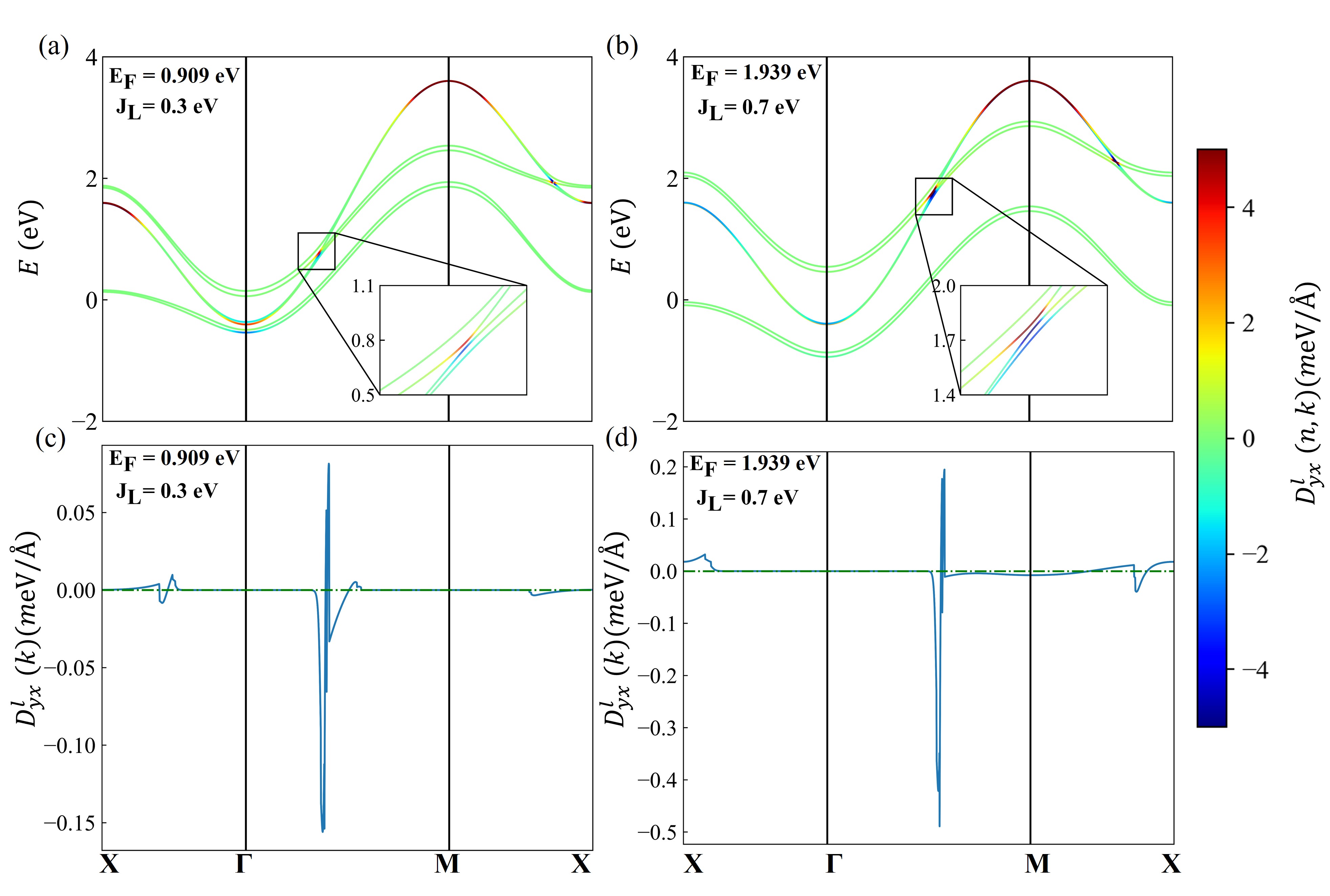} 
\caption{{\bf Orbital DMI with varying $J_L$}. The band-projected orbital DMI along the high symmetry path in the Brillouin zone at (a) $E_F=0.909$\,eV for $J_L=0.3$\,eV and (b) $E_F=1.939$\,eV for $J_L=0.7$\,eV. The color marks the value of orbital DMI projected on each band. Clearly, the band avoided crossing region contributes primarily to the  orbital DMI. (c-d) The $\mathbf{k}$-resolved orbital DMI along the high symmetry path in the Brillouin zone corresponding to the case in (a) and (b), respectively.} 
\label{Fig.5} 
\end{figure}

In Fig.~4(b) we show the variation of the orbital DMI $D_{y x}^{\mathrm{L}}$ with the inversion symmetry breaking strength as given by parameter $\gamma$. As mentioned above, the strength of inversion symmetry breaking directly influences the ``intensity" of orbital hybridization, ultimately resulting in the formation of chiral orbital textures. We observe that the orbital DMI monotonically increases with the gradual increase of  $\gamma$, and zero $\gamma$ results in vanishing orbital DMI.
This highlights the decisive role of inversion symmetry breaking as a dominant factor for orbital DMI. The presence of ISB, along with the induced orbital hybridization and chiral orbital textures, governs the existence of orbital DMI, consistent with previous studies on the orbital Hall effect~\cite{4}.

The analysis of  band-projected and $\mathbf{k}$-resolved orbital DMI for $\gamma=0.005$\,eV and $\gamma=0.015$\,eV (results for other values of $\gamma$ can be found in the supplementary section S5~\cite{63}), shown in Figs.~6(a) and 6(b), reveals that the position of the orbital hybridization points, which contribute predominantly to the orbital DMI, remains unaffected as $\gamma$ undergoes variation within the considered range of values. Consequently, there is no discernible shift of the peak position in the orbital DMI curve, as evident from Fig.~4(b). As shown in Figs.~6(c) and 6(d), it is also directly evident that the increase in the degree of  inversion symmetry breaking results in an amplification of orbital hybridization at the respective hybridization points, as reflected in the degree of orbital mixing~\cite{64}. This leads to an enhancement of the local orbital DMI, and an increase in the peak height of the orbital DMI curve in Fig.~4(b).

At last, we present in Figs.~4(c) and 4(d) the dependence of $D_{y x}^{\mathrm{L}}$ on the angle of the orbital exchange field $\theta$.
Specifically, the range of $\theta$ from $0^{\circ}$ to $45^{\circ}$ is depicted in Fig.~4(c), while the range from $45^{\circ}$ to $90^{\circ}$ is shown in Fig.~4(d). We can clearly observe that for $\theta$ less than $45^{\circ}, D_{y x}^{\mathrm{L}}$ gradually decreases with increasing $\theta$ at around $E_F=2.7 \mathrm{eV}$, while in other energy regions the DMI values stay stable. On the other hand, for $\theta$ values above $45^{\circ}$, the situation is reversed, and it is the DMI between 0 and 1\,eV which is increasing rapidly with increasing $\theta$ on the background of relative stability at other energies.
Such behavior is a manifestation of strong anisotropy of orbital DMI with respect to the direction of the orbital exchange field. To understand this effect better, we plot the band-projected and $\mathbf{k}$-resolved orbital DMI for two orientations of the orbital exchange field: the out-of-plane direction ($\theta=0^{\circ}$), and the in-plane direction ($\theta=90^{\circ}$). As shown in Fig.~7, drastic changes in the band structure are observed as the orbital exchange field direction varies. As a result, the position of the band avoided crossing region that mainly contributes to the orbital DMI experiences alteration. Simultaneously, the degree of orbital hybridization also exhibits pronounced variation, manifested as changes in the magnitude of orbital DMI at the hybridization positions. These factors combined lead to modifications in both the peak position and peak height of the orbital DMI curve as the orbital exchange field direction transitions from the out-of-plane to the in-plane configuration, showcasing a strong anisotropy, as depicted in Figs.~4 (c) and 4(d).

\begin{figure}[t!] 
\centering 
\includegraphics[width=0.48\textwidth]{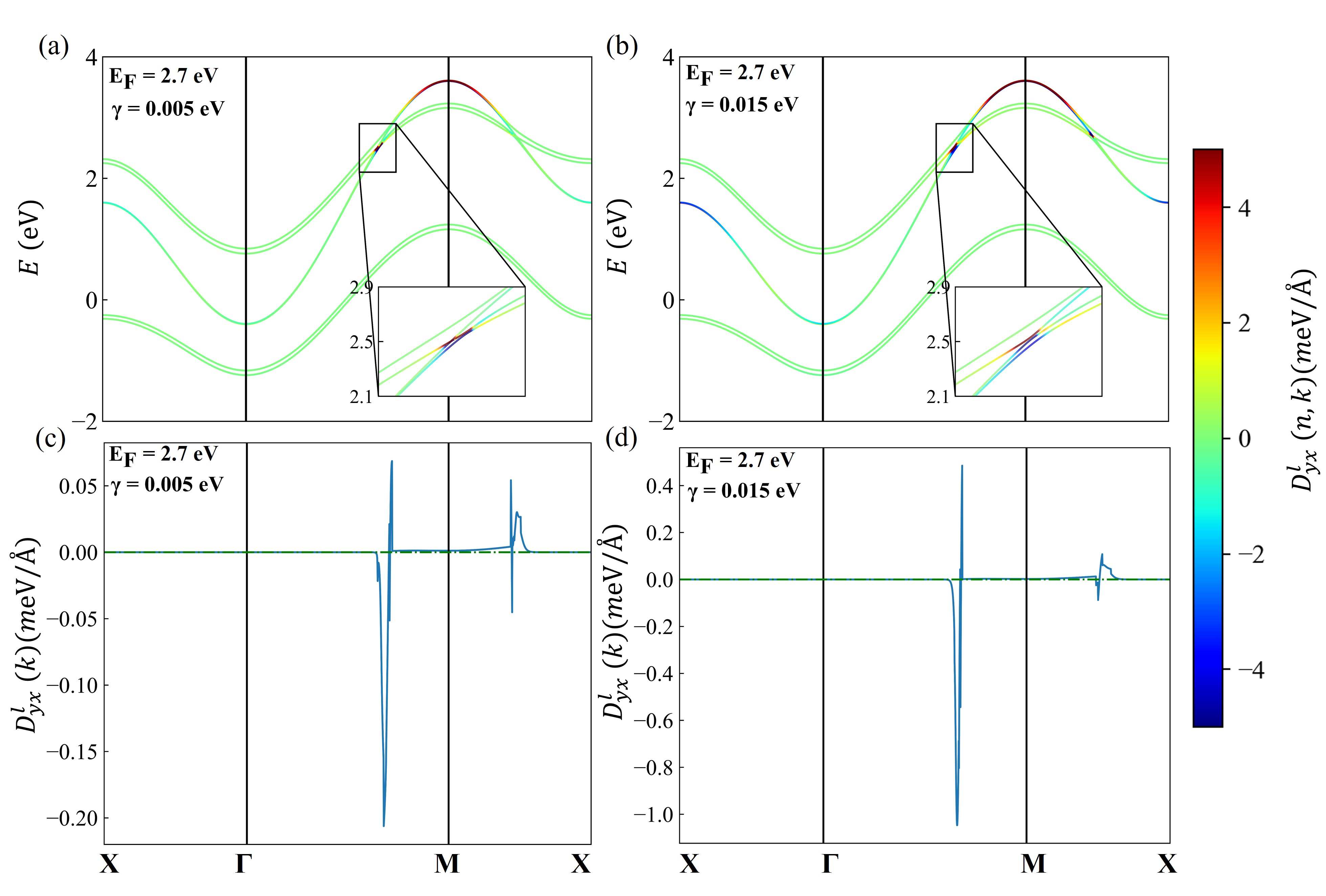} 
\caption{{\bf Orbital DMI with varying $\gamma$}. The band-projected orbital DMI along the high symmetry path in the Brillouin zone at $E_F=2.7$\,eV for (a) $\gamma=0.005$\,eV and (b)  $\gamma=0.015$\,eV. The color marks the value of orbital DMI projected on each band. Clearly, the band avoided crossing region contributes primarily to the orbital DMI. (c-d) The $\mathbf{k}$-resolved orbital DMI along the high symmetry path in the Brillouin zone corresponding to the case in (a) and (b), respectively.} 
\label{Fig.6} 
\end{figure}

\begin{figure}[t!] 
\centering 
\includegraphics[width=0.48\textwidth]{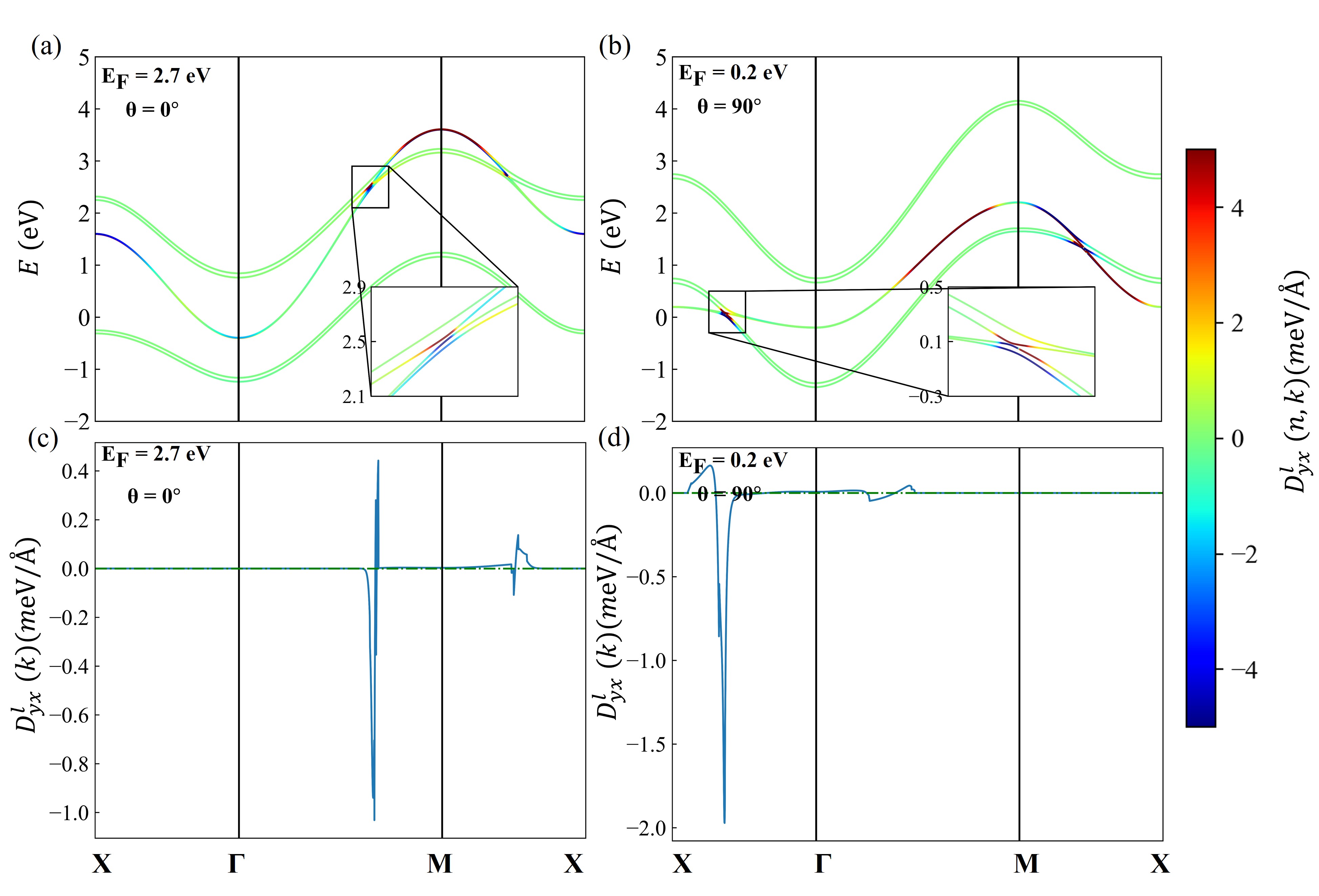} 
\caption{{\bf Orbital DMI with varying $\theta$}. The band-projected orbital DMI along the high symmetry path in the Brillouin zone at (a) $E_F=2.7$\,eV for $\theta=0^{\circ}$ and (b) $E_F=0.2$\,eV for $\theta=90^{\circ}$. The color marks the value of orbital DMI projected on each band. (c-d) The $\mathbf{k}$-resolved orbital DMI along the high symmetry path in the Brillouin zone corresponding to the case in (a) and (b), respectively.} 
\label{Fig.7} 
\end{figure}

\subsection{Relation between orbital DMI and spin DMI}

Finally, we consider the case when in addition to the orbital exchange and SOC the spin exchange interaction as given by $H_{\mathrm{exc}}^{\mathrm{S}}$ is also present in the Hamiltonian. We keep the values of 
$J_L$ and $J_S$ at 1\,eV, while keeping the out-of-plane directions of the spin and orbital exchange fields, and compute both spin and orbital DMI by following the Berry phase theory outlined above.
We focus specifically on the dependence of the spin and orbital DMI on the SOC strength, presenting the results in  Fig.~8. First, we consider the regime of small $\lambda_{\rm soc}$, Figs.~8(a-b). We observe that in this regime the orbital DMI exhibits minimal variations, whereas the spin DMI consistently increases with increasing the SOC strength, vanishing identically without spin-orbit interaction. More importantly, we observe that for the energies above +1.5\,eV the qualitative behavior of the spin and orbital DMI is quite similar in that both quantities develop large peaks at about 2.2 and 3.3\,eV, albeit of opposite sign in the case of spin.

\begin{figure}[t!] 
\centering 
\includegraphics[width=0.5\textwidth]{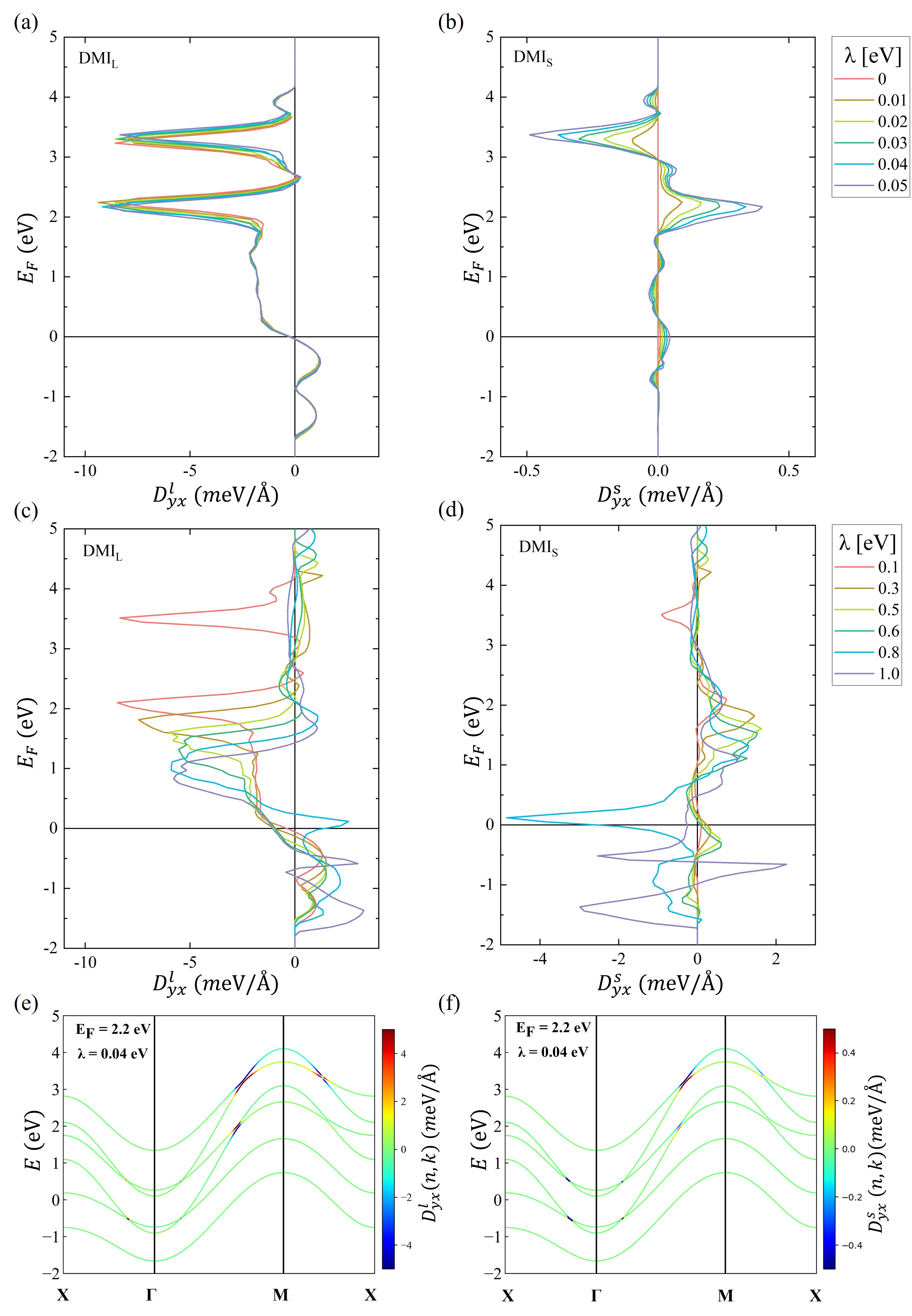}
\caption{{\bf Variation of spin and orbital DMI with SOC}. (a) The orbital DMI $\left(D_{y x}^{\mathrm{L}}\right)$ and (b) spin DMI $\left(D_{y x}^{\mathrm{S}}\right)$ as a function of the Fermi energy $\left(E_F\right)$ for various SOC strengths $\lambda$ in small SOC regime. (c) $D_{y x}^{\mathrm{L}}$ and (d) $D_{y x}^{\mathrm{S}}$ in large SOC regime. (e-f) The band-resolved contributions to the orbital (e) and spin (f) DMI for spin-orbit strength of $\lambda_{\rm soc}=0.04$\,eV at $E_F=2.2$\,eV.} 
\label{Fig.8} 
\end{figure}

To understand the origin of this correlation, we scrutinize the band-resolved contributions to the spin and orbital DMI at the value of $\lambda_{\rm soc}=0.04$\,eV and $E_F=2.2$\,eV, shown in Figs.~8(e-f). We observe that in the discussed region of energy both falvors of DMI come from the same anticrossings in the electronic structure along $\Gamma$M. These anticrossings are nothing else but the ISB-driven hybridization points between orbitally-different bands $-$  discussed in depth above and shifted in energy by spin exchange splitting $-$ which induce orbital DMI without SOC. We thus come to a fundamental conclusion that the orbital DMI is the effect which is parent to spin DMI. As in the case of a relation between the orbital Hall effect and spin Hall effect~\cite{4}, the spin DMI is ``pulled" by the orbital DMI via SOC. In the case when DMI-driving hybridizations are 
well-defined in energy and $\mathbf{k}$-space, such as e.g. for energies above $+1.5$\,eV, the behavior of spin and orbital DMI can be very similar but not necessarily identical: for example, in the case considered here, the switch in sign correlation  between $D_{y x}^{\mathrm{L}}$ and $D_{y x}^{\mathrm{S}}$ at $+2.2$ and $+3.3$\,eV can be explained by the opposite sign of the spin-orbit correlation $\langle LS \rangle$ of participating bands~\cite{4} at these energies, similarly to the case of Hall effects.

As a result, when at a given energy the DMI contributions are small and spread over several bands with different orbital character filling larger areas of $\mathbf{k}$-space, as it is the case below $+1$\,eV, Figs.~8(a-b), the correlation between $D_{y x}^{\mathrm{L}}$ and $D_{y x}^{\mathrm{S}}$ is much less pronounced, or not at all present.
This is ultimately the reason why it is difficult to observe a direct relation between   $D_{y x}^{\mathrm{L}}$ and $D_{y x}^{\mathrm{S}}$ for the case of larger SOC strength,
shown in Figs. 8(c-d). In the latter case, the strongly modified by SOC bands develop larger areas in $\mathbf{k}$-space where the effect of SOC on the DMI is active. This is also consistent with the saturation in the values of the spin DMI with increasing SOC towards the values comparable to the strength of exchange interaction. 
It is especially worth mentioning that when the strength of orbital exchange interaction $J_L$ is similar to  the magnitude of spin exchange $J_S$, the values of the orbital DMI  on average exceed those of the spin DMI by an order of magnitude. This underlines the fact that the orbital angular momentum in solids is much more prone to the effects of chiralization when compared to spin, due to its stronger coupling to the lattice and qualitatively different energetics of orbital dynamics relying on crystal field structure rather than spin-orbit interaction.

\section{CONCLUSION}
In summary, we theoretically demonstrate that the  chiral exchange interaction among orbital moments, closely resembling the spin Dzyaloshinskii-Moriya interaction, can arise in a two-dimensional square lattice with $t_{2 g}$ orbitals. We find that orbital hybridization, induced by inversion symmetry breaking, is crucial for generating orbital DMI. Furthermore, the position and strength of orbital hybridization have a crucial impact on both the magnitude and distribution of orbital DMI. We also argue that in the vicinity of isolated Fermi surface features the orbital DMI is closely correlated with the behavior of the orbital mixed Berry curvature. Moreover, the strength of orbital exchange interaction exerts a decisive impact on the magnitude of orbital DMI. In situations where the strength of the orbital exchange interaction is comparable to that of the spin exchange interaction, the orbital DMI can exceed the spin DMI by an order of magnitude. Importantly, consistent with other orbital phenomena, the orbital DMI can also emerge in the absence of spin-orbit interaction, subsequently inducing the spin DMI through the effect of SOC. 
While with our work we provide first basic insights into the physics and properties of orbital DMI, further experimental verification is required to validate the existence and control of orbital DMI. On the other hand, more effort is needed to suggest specific material candidates where the formation of chiral orbital textures could be experimentally observed.

\begin{acknowledgments}
This work was supported by the Deutsche Forschungsgemeinschaft (DFG, German Research Foundation) - TRR
173/2 - 268565370 (project A11), and TRR 288 – 422213477
(project B06). 
The authors would also like to thank the supports by the projects from National Natural Science Foundation of China (No.61627813, 62204018 and 61571023), the Beijing Municipal Science and Technology Project under Grant Z201100004220002, the National Key Technology Program of China 2017ZX01032101, the Program of Introducing Talents of Discipline to Universities in China (No. B16001), the VR innovation platform from Qingdao Science and Technology Commission. 
\end{acknowledgments}

\bibliography{Ref}

\end{document}